\documentclass[3p,times]{elsarticle}

\usepackage{ecrc}
\usepackage{hyperref}
\usepackage{slashed}

\usepackage{epstopdf}

\volume{00}

\firstpage{1}

\journalname{Nuclear Physics A}

\runauth{N.~Armesto et al.}

\jid{nupha}

\jnltitlelogo{Nuclear Physics A}

\usepackage{graphicx}
\usepackage{amsmath,amssymb}

\usepackage{lineno}

\newcommand{\sqrtsNN}{\sqrt{s_{\rm \scriptscriptstyle NN}}}

\begin{document}

\begin{frontmatter}



\title{Nuclear collisions at the Future Circular Collider}



\author[label1]{N.~Armesto}
\ead{nestor.armesto@usc.es}
\author[label2]{A.~Dainese}
\author[label3]{D.~d'Enterria}
\author[label4]{S.~Masciocchi}
\author[label5]{C.~Roland}
\author[label1]{\\C.A.~Salgado}
\author[label7]{M.~van Leeuwen}
\author[label3]{U.A. Wiedemann}

\address[label1]{Departamento de F\'isica de Part\'iculas and IGFAE, Universidade de Santiago de Compostela, 15706 Santiago de Compostela, Galicia-Spain}
\address[label2]{INFN - Sezione di Padova, 35131 Padova, Italy}
\address[label3]{Physics Department, CERN, CH-1211 Gen\'eve 23, Switzerland}
\address[label4]{EMMI and GSI Helmholtzzentrum fuer Schwerionenforschung GmbH, Darmstadt, Germany}
\address[label5]{Massachusetts Institute of Technology, Cambridge, MA 02139-4307, USA}
\address[label7]{Nikhef, National Institute for Subatomic Physics, Amsterdam and Institute
for Subatomic Physics of Utrecht University, Utrecht, Netherlands}

\begin{abstract}
The Future Circular Collider is a new proposed collider at CERN with centre-of-mass energies around 100 TeV in the pp mode. Ongoing studies aim at assessing its physics potential and  technical feasibility. Here we focus on updates in physics opportunities accessible in pA and AA collisions not covered in previous Quark Matter contributions, including Quark-Gluon Plasma and gluon saturation studies, novel hard probes of QCD matter, and photon-induced collisions.
\end{abstract}


\end{frontmatter}



\section{Introduction}
\label{intro}

In 2014, an international design study for a new accelerator at CERN, the Future Circular Collider (FCC)~\cite{FCCweb}, started with the goal of preparing a Conceptual Design Report for the next European Strategy for Particle Physics to be discussed in 2018. As an intermediate step, a physics white paper is to be produced for spring 2016. The study includes electron-positron (ee), electron-hadron (he) and hadron-hadron (hh) modes. A centre-of-mass energy $\sqrt s$ of the order of 100~TeV is aimed for pp collisions which, using the relation $\sqrtsNN= \sqrt {s_{\rm pp}} \sqrt{Z_1 Z_2 / A_1 A_2}$, would correspond to $\sqrtsNN \sim 39$~TeV for PbPb ($Z=82$, $A=208$) and 63~TeV for pPb collisions, and rapidities of the  proton and Pb beams $\sim 11.6$ and 10.6 respectively.
Studies for a new 80-100~km tunnel in the Geneva area required for such facility, detector design and cost evaluation  are ongoing, with 
a target operation start around 2035-2040.

Present estimates of the peak and integrated luminosities for PbPb collisions are $13\cdot 10^{27}$~cm$^{-2}$s$^{-1}$ and 8~nb$^{-1}$ per month of running respectively~\cite{schaumann,JohnWashington}, the latter being about eight times larger than
the current projection for the future LHC runs~\cite{rliup}. The corresponding numbers for pPb collisions are $3.5\cdot 10^{30}$~cm$^{-2}$s$^{-1}$ and 1800~nb$^{-1}$ respectively~\cite{schaumann,JohnWashington}.

The increased centre-of-mass energy and luminosity with respect to the LHC open up new opportunities. The topics that are under current exploration are:\\
$\bullet$  Quark-Gluon Plasma (QGP) studies with soft observables (section~\ref{sec:QGP});\\
$\bullet$ novel possibilities with hard probes (section~\ref{sec:hardprobes});\\
$\bullet$ high-density QCD and gluon saturation in the initial state of heavy-ion collisions (section~\ref{sec:others});\\
$\bullet$ photon-induced collisions (section~\ref{sec:others}).\\
\indent
In these proceedings, we further support the physics case for nuclear collisions at the FCC with arguments not yet covered in previous Quark Matter contributions \cite{Armesto:2014iaa}.
Detector considerations are not discussed, because
no specific studies have been carried out yet. The present view is to explore the feasibility of a heavy-ion programme with 
general-purpose detectors.
More details can be found in~\cite{Armesto:2014iaa,AndreaWashington,FCCAAworkshops}. Note that the physics case for nuclear collisions at  $\sqrtsNN$ of the order of several tens of TeV in a future hadron accelerator in China has been elaborated in~\cite{Chang:2015hqa}.

\vspace{-3mm}

\section{Study of the hot and dense Quark-Gluon Plasma with soft observables}
\label{sec:QGP}

The QGP medium produced in PbPb collisions at $\sqrtsNN=39$~TeV is expected to have larger volume, lifetime, energy density and temperature 
than at the top LHC energy. Estimations of the corresponding charged particle multiplicity,  transverse energy, volume and lifetime, based on extrapolations from lower energies, can be found in  \cite{Armesto:2014iaa}, together with those for temperature and energy density built on simple parametric ideas. The charged particle multiplicity is expected to increase by 
about a factor of two from top LHC to FCC energy. The volume and lifetime of the system are foreseen to increase by a factor of two and by 20\%, 
respectively. The energy density will increase by a factor of two, reaching for example a value of 35~GeV/fm$^3$ at 1~fm/$c$. The increase of temperature at  a given time is modest, but the formation time of the system 
(i.e. the initial time of the QGP evolution) can be expected to be smaller at the FCC than at the LHC, where it is usually taken as $\tau_0\sim 0.1$~fm/$c$. If the formation time 
is significantly lower than 0.1~fm/$c$, the initial temperature could be as large as $T_0\approx800$~MeV. 

\begin{figure}[htb]
\begin{center}
\includegraphics[width=0.49\textwidth]{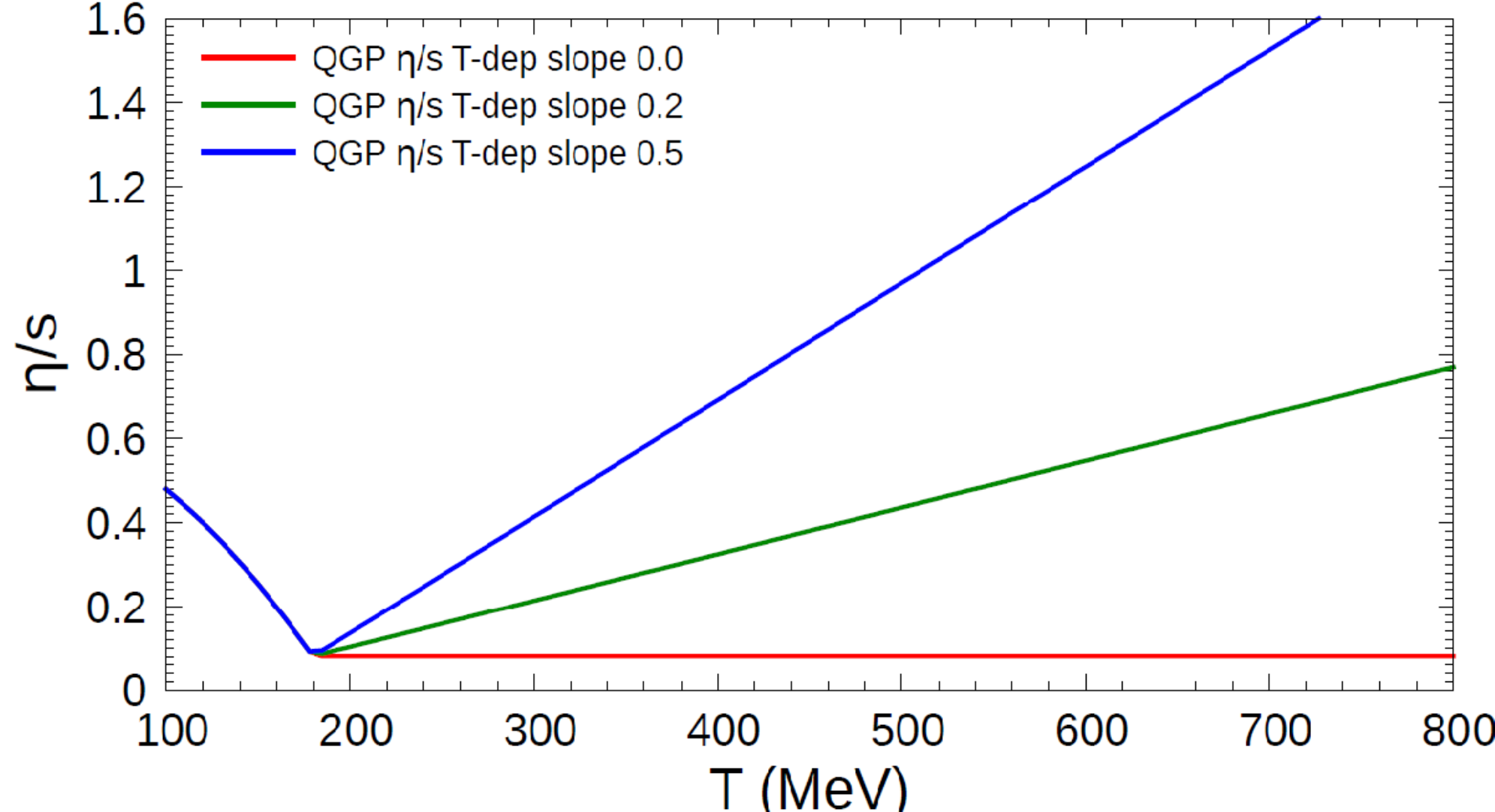}
\hfill
\includegraphics[width=0.46\textwidth]{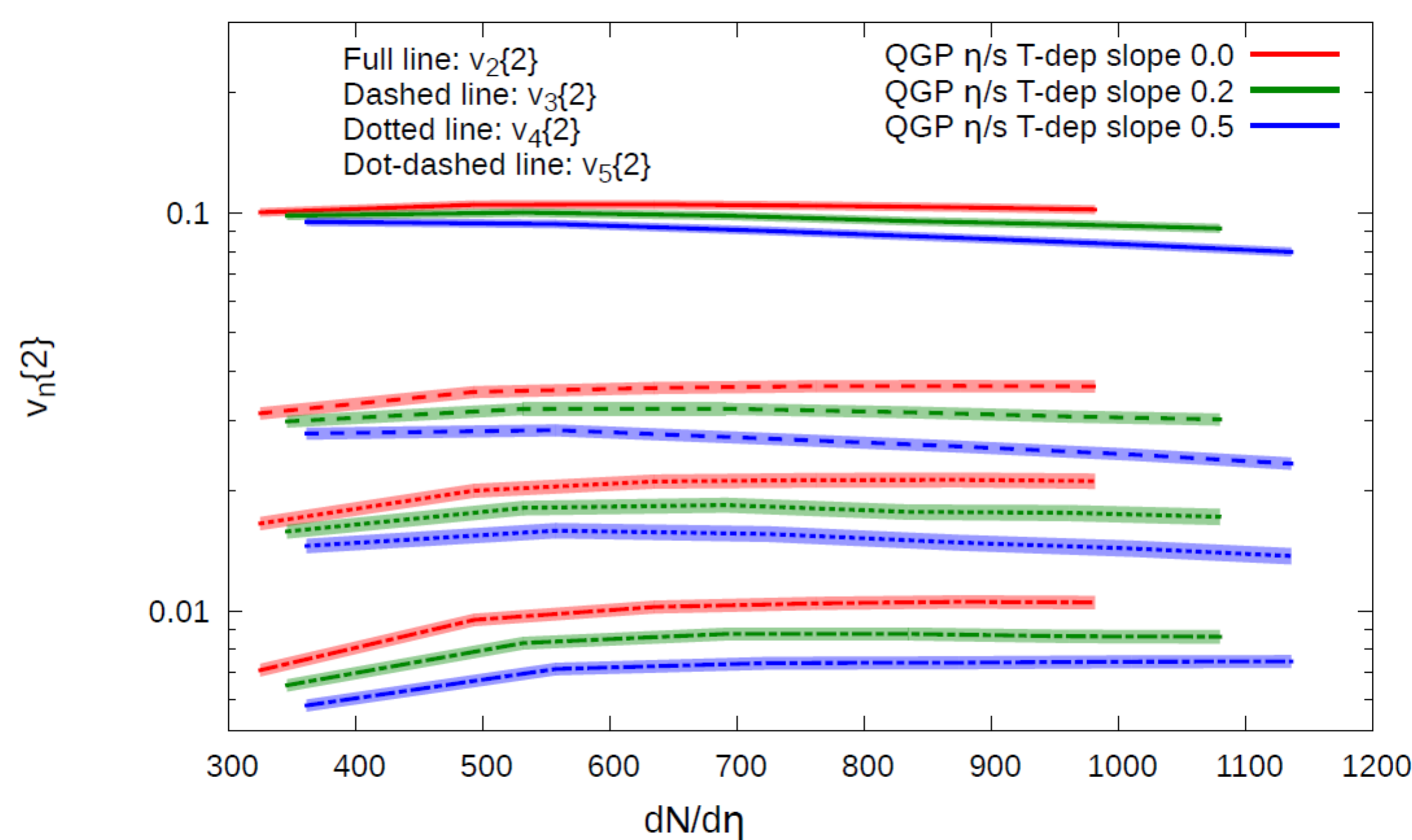}
\caption{Left: Parametrisations for the evolution of the ratio $\eta/s$ versus temperature.
Right: Results for different $v_n\{2\}$ versus multiplicity from viscous hydrodynamics calculations for different temperature dependencies of $\eta/s$ shown on the left. Taken from \cite{denicol}.}
\label{fig:flow}
\end{center}
\end{figure}

The higher initial temperature and density will also lead to a (moderate) increase in the collective flow and a larger final volume. In addition, the two-fold larger multiplicity
may open up the possibility to carry out flow measurements on an event-by-event basis and to become sensitive to dependencies of transport coefficients that are very difficult to address at the LHC. For example, the different azimuthal coefficients $v_n$ are sensitive to the dependence of the ratio shear viscosity over energy density, $\eta/s$, on the temperature, and this dependence becomes stronger with increasing multiplicity and for higher harmonics. This is illustrated in Fig.~\ref{fig:flow}, taken from \cite{denicol}.

\begin{figure}[htb]
\begin{center}
\vskip-0.4cm
\centering
\begin{minipage}{0.5\textwidth}
\centering
\includegraphics[width=0.68\textwidth,angle=270]{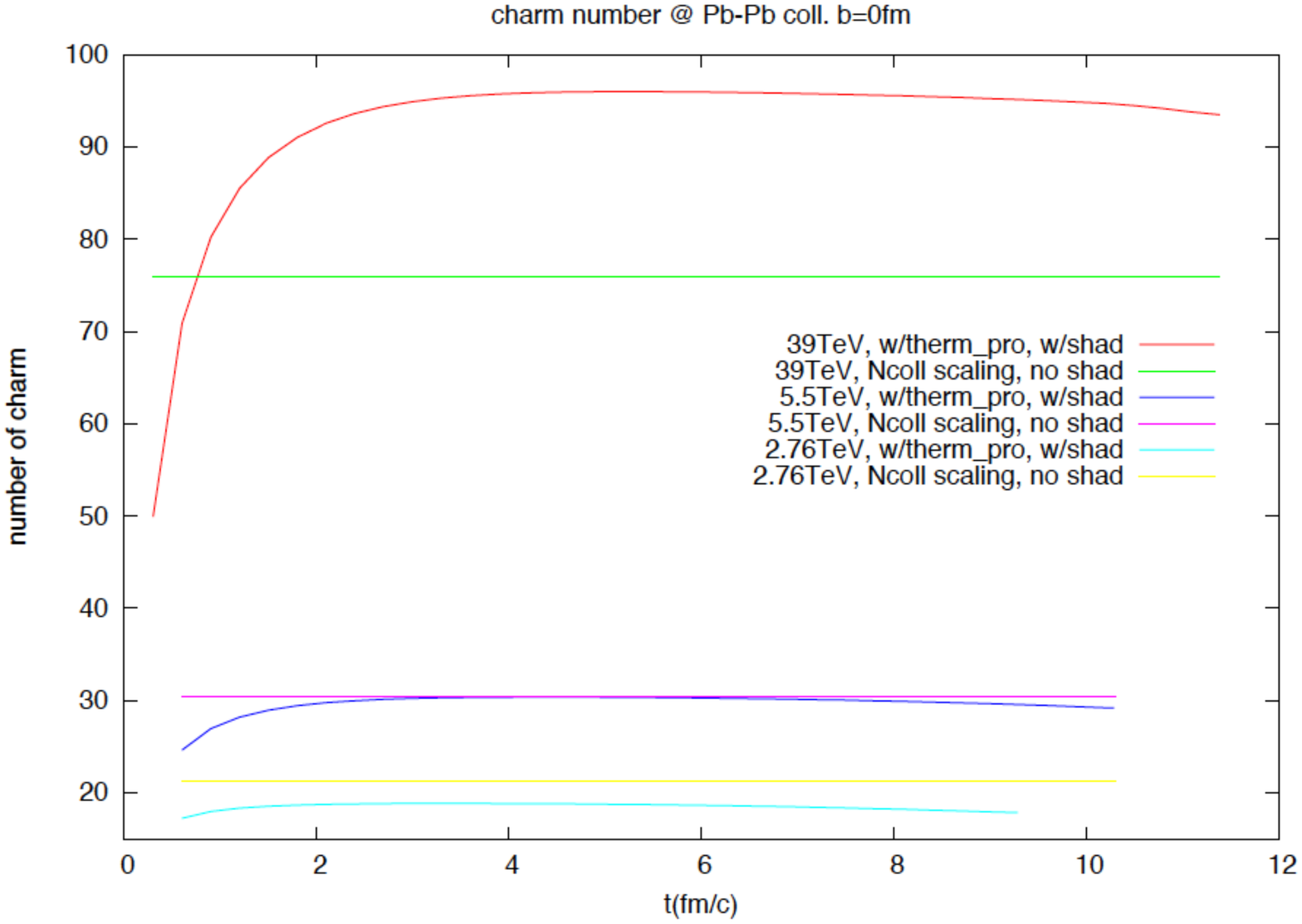}
\end{minipage}%
\begin{minipage}{0.5\textwidth}
\centering
\includegraphics[width=0.78\textwidth]{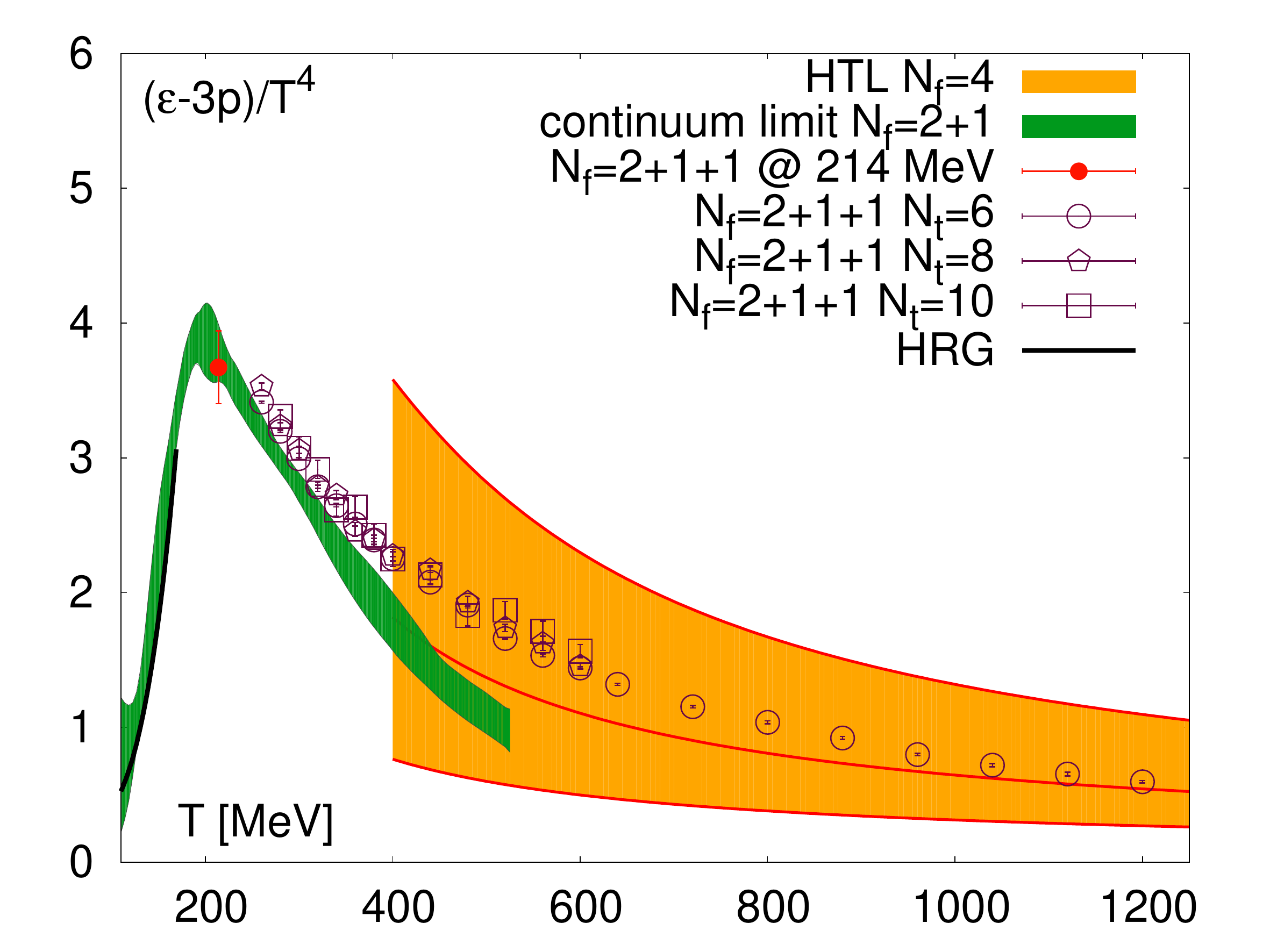}
\end{minipage}
\vskip -0.4cm
\caption{Left: Time-evolution of the charm-anticharm yield (per unit of rapidity at central rapidity) for central PbPb collisions 
at $\sqrtsNN=2.76$ (lower lines), 5.5 (lines in the middle) and 39 (upper lines) TeV \cite{zhou}. Results with shadowing and thermal production, and without both, are shown for each energy.
Right: Trace anomaly computed in lattice QCD with different numbers of flavours, and in perturbative QCD, taken from \cite{Borsanyi:2014rza}.}
\label{fig:charm}
\end{center}
\end{figure}

\vskip -0.4cm
Another interesting consequence of the increase in the temperature of the system could be a sizeable production of secondary charm-anticharm ($\rm c\overline c$) pairs from partonic
interactions during the hydrodynamical evolution of the system.
Calculations for FCC energy indicate that
 this secondary production can be of the same order of magnitude as the initial charm production in hard scattering processes,
overcoming the shadowing of nuclear parton densities (nPDFs) that is expected to be larger at the FCC than at the LHC. Moreover, it is very sensitive to the initial temperature and temperature evolution of the QGP. As an example, Fig.~\ref{fig:charm} left shows the time evolution of the 
number of $\rm c\overline c$ pairs per unit of rapidity at central rapidity (the value at initial time represents the production from hard scattering), computed at next-to-leading order using rate equations within a 2+1 hydrodynamical evolution \cite{zhou}. The secondary charm production would yield an enhancement of charmed hadron 
production in the very-low-$p_{\rm T}$ region, with respect to the expectation from binary scaling of the production in pp collisions. This enhancement potentially provides a handle on the
temperature of the QGP. In addition, the abundance of charm quarks in the QGP is expected to have an effect on its equation of state: lattice QCD calculations show a sizeable 
increase of $P/T^4$ (which is proportional to the number of degrees of freedom) and of the trace anomaly when the charm quark is included and the system temperature is larger than about 400~MeV~\cite{Borsanyi:2014rza}, see Fig.~\ref{fig:charm} right.

\vskip -3mm
\section{Novel possibilities with hard probes}
\label{sec:hardprobes}

\begin{figure}[htb]
\vskip -0.4cm
\begin{center}
\includegraphics[width=0.42\textwidth]{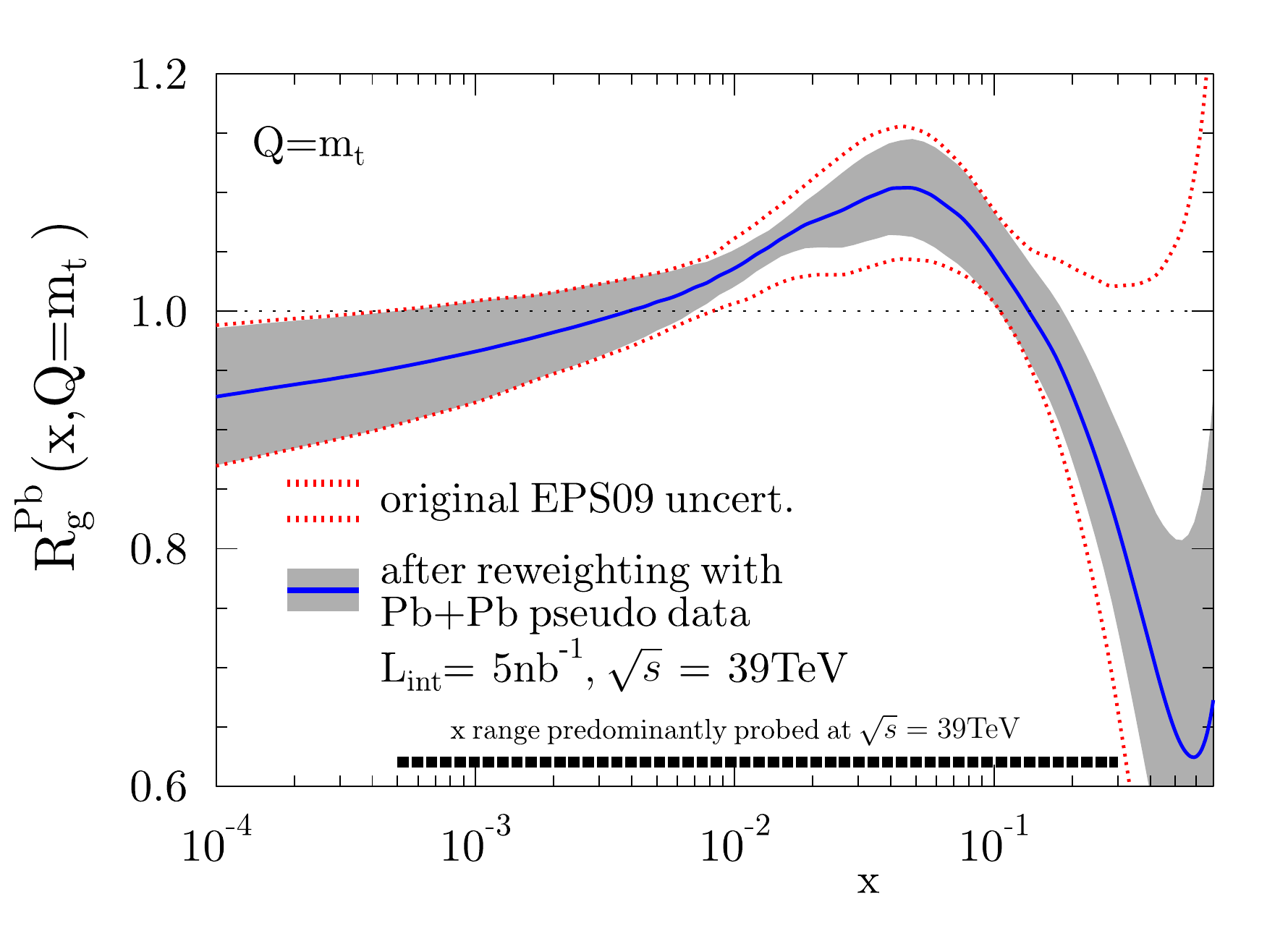}
\hfill
\includegraphics[width=0.42\textwidth]{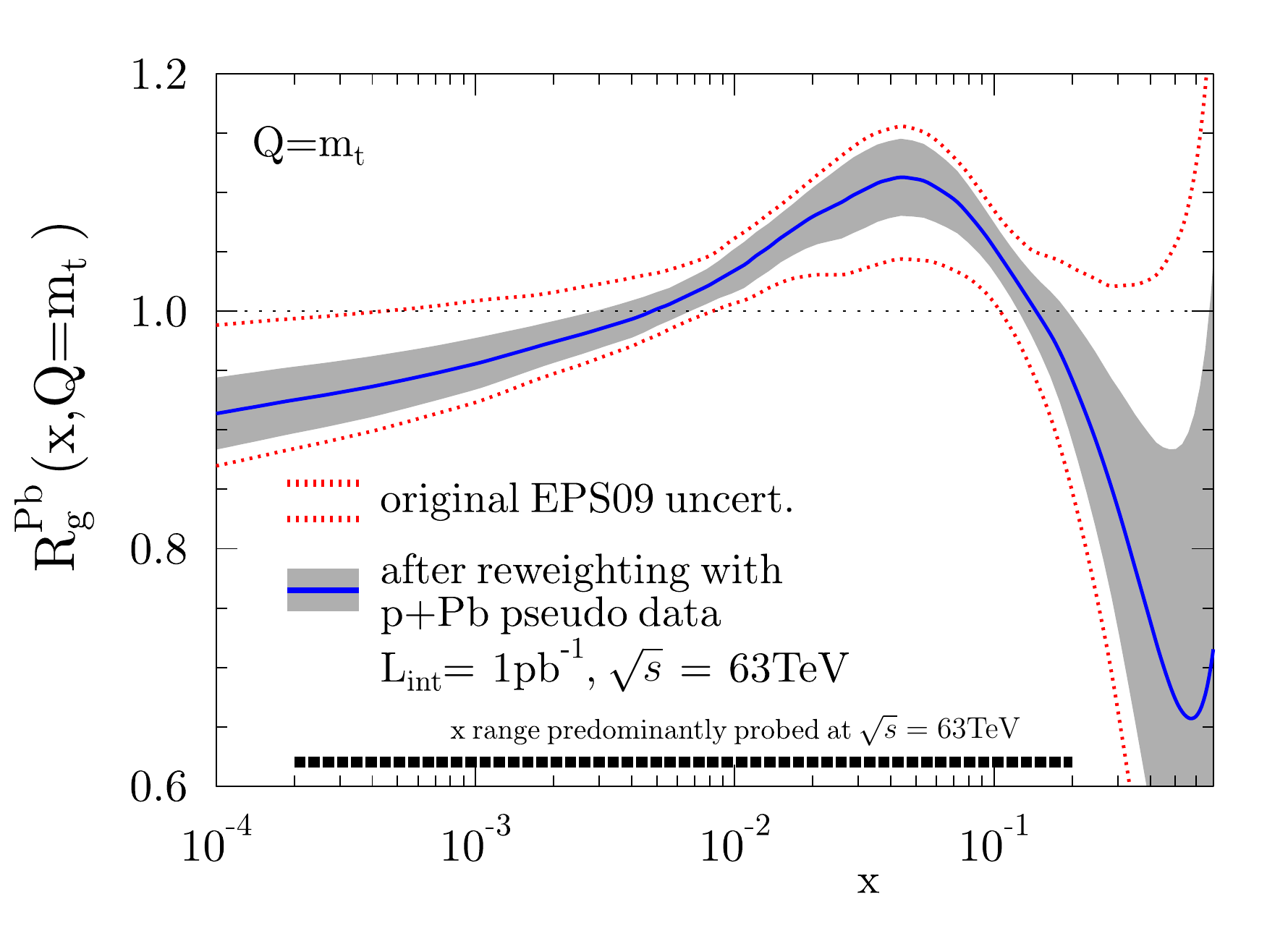}
\vskip -0.4cm
\caption{Modification of the uncertainties in the gluon densities in the EPS09 framework, resulting from the inclusion of top pseudodata using the Hessian reweighting method, for PbPb (left) and pPb (right) collisions at the FCC. Taken from \cite{d'Enterria:2015jna}.}
\label{fig:gluons}
\end{center}
\end{figure}

The larger energy and luminosities will make new, rarer, hard probes available. 
E.g. the cross sections for 
top, $Z$+$1\,{\rm jet}\,(p_{\rm T}>50~{\rm GeV}/c)$, bottom and $Z$ increase by factors 80, 20, 8 and 7, respectively, from top LHC to FCC energy \cite{Armesto:2014iaa}.
A novel aspect that may be explored at the FCC is top production. Yields per year of PbPb and pPb running of tens of thousands of $t\,\bar t\to b\,\bar b \,l \,\bar l \,\nu \,\bar \nu$, and of thousands of $t \, W\to b \,l\, \bar l\, \nu\, \bar \nu$, are expected after typical efficiency and acceptance cuts \cite{d'Enterria:2015jna}. Two  possibilities are currently being explored: On the one hand, measurements of the lepton spectra from top decays will reduce the current uncertainties in nPDFs \cite{d'Enterria:2015jna}, see Fig. \ref{fig:gluons} where a sizeable reduction of the uncertainty band is obtained, particularly for pPb collisions (for the LHC the constraining power is far smaller).

\begin{figure}[htb]
\begin{center}
\includegraphics[width=0.47\textwidth]{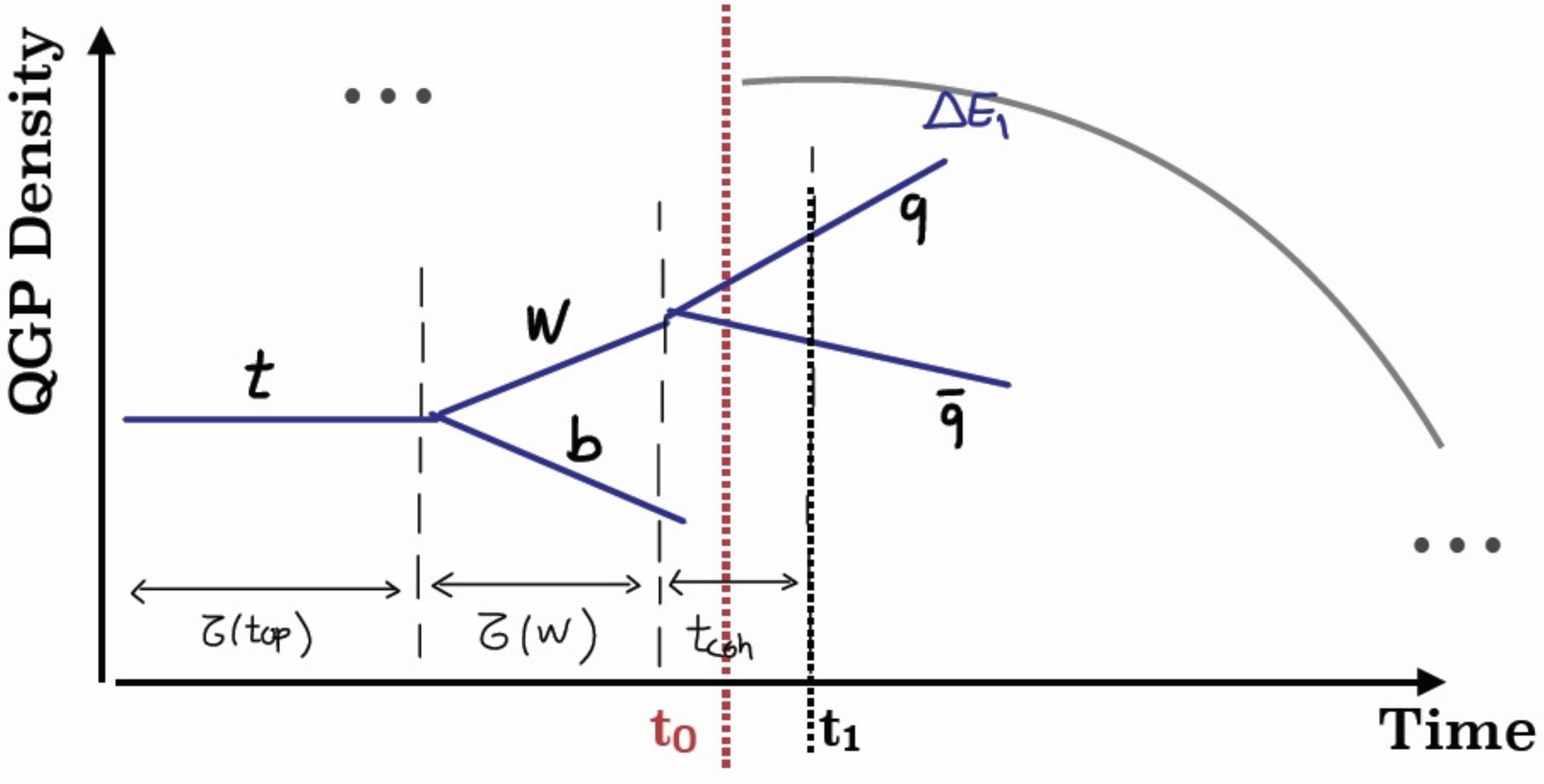}
\hfill
\includegraphics[width=0.47\textwidth]{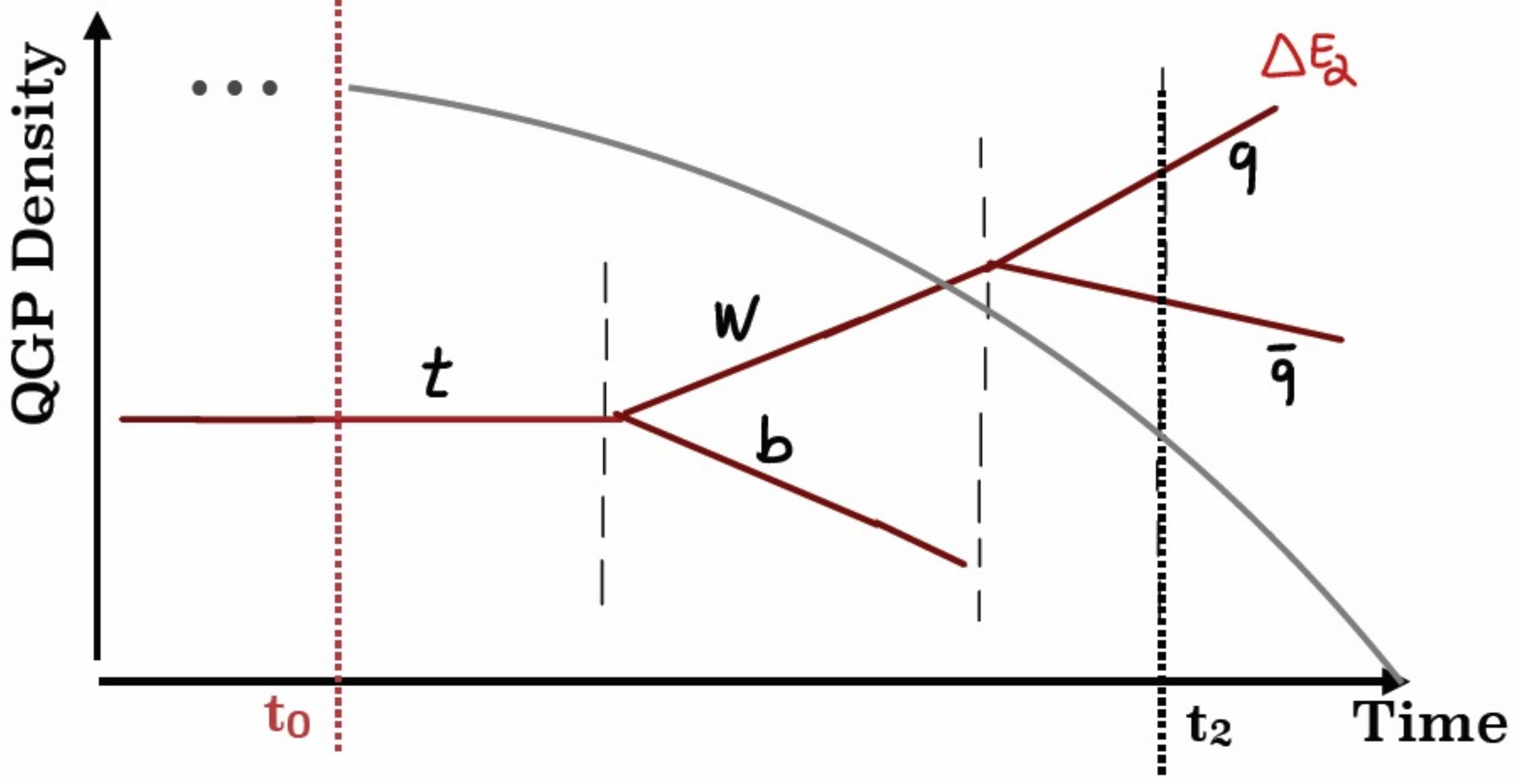}
\caption{Illustration of the different time scales for top and $W$ decay and for the decoherence time of the colourless $q\bar q$ pair from $W$ decay, compared with the formation $t_0$ and freeze-out times of the medium, for slow (left) and fast (right) tops. Taken from \cite{liliana}.}
\label{fig:tops}
\end{center}
\end{figure}

\vskip -0.5cm
In addition, identified tops (in the channel  $t\,\bar t\to b\, \bar b\, +l+2\ {\rm jets}+\slashed{E}_T$) in PbPb collisions may provide information about the space-time evolution of the QGP \cite{liliana}. As illustrated in Fig.~\ref{fig:tops}, tops decay into $W+b$ after some decay time; then, the process $W\to  2\ {\rm jets}$ produces a colourless $q\bar q$ pair that, depending on the boost of the top, may be born inside or outside the medium. Furthermore, medium effects on the formed colourless antenna would determine how fast its colour coherence is broken, and thus alter the pattern of gluon radiation off the two-jet system. For example, the energy loss of any of the legs on the antenna would be smaller for fast (Fig.~\ref{fig:tops} right) than for slow tops (Fig.~\ref{fig:tops} left).

\vskip -3mm
\section{Photon-induced collisions and connections with other fields}
\label{sec:others}

\begin{figure}[htb]
\begin{center}
\includegraphics[width=0.44\textwidth]{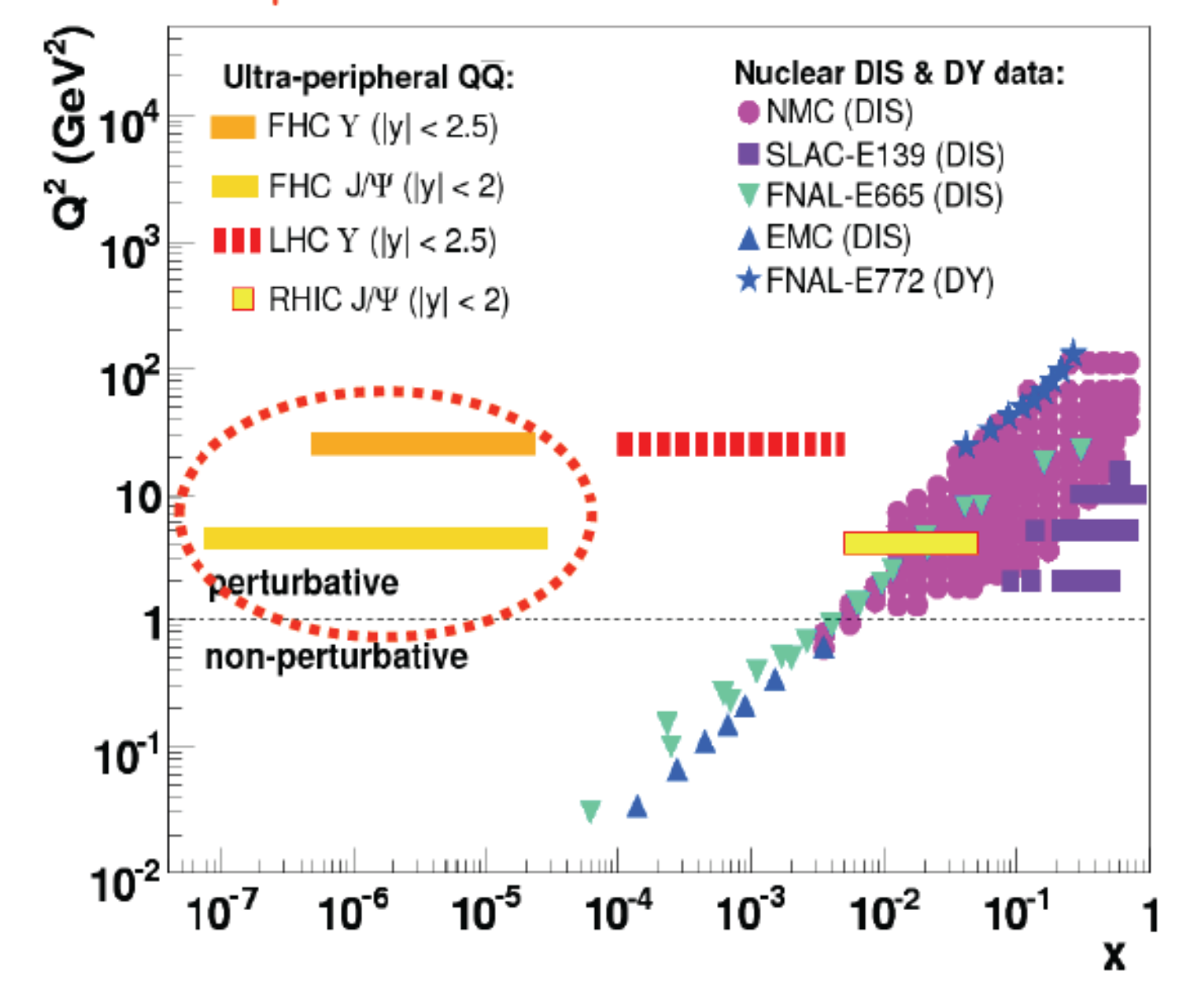}
\hfill
\includegraphics[width=0.4\textwidth]{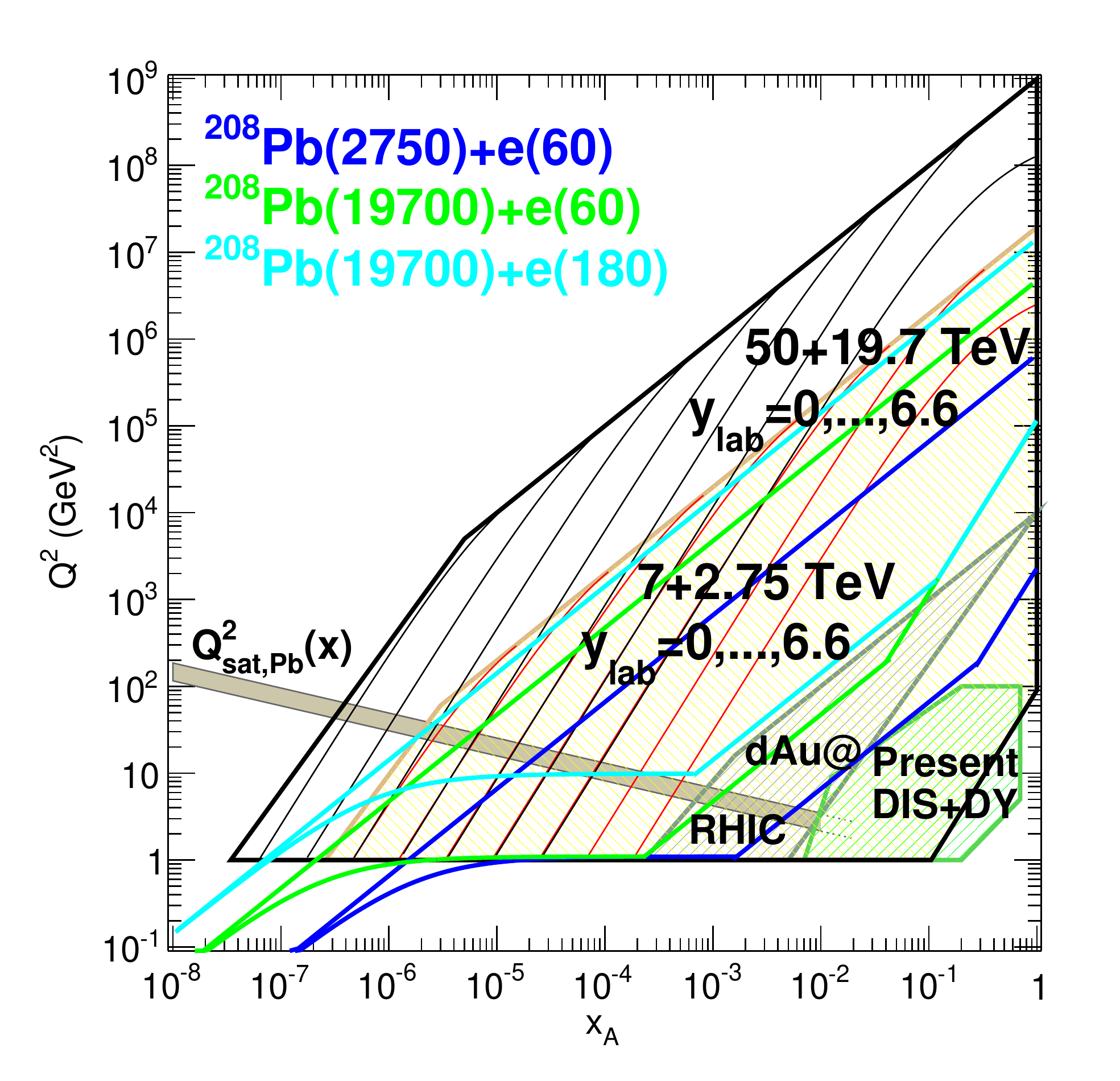}
\vskip -0.4cm
\caption{Left: $x-Q^2$ plane explored with presently available nuclear DIS and Drell-Yan data, and in UPCs at RHIC, the LHC and the FCC using exclusive $J/\psi$ and $\Upsilon$ production, taken from \cite{dEnterria}. Right: $x-Q^2$ plane explored with presently available nuclear DIS and Drell-Yan data, in dAu collisions at RHIC and pPb collisions at the LHC and at the FCC, and for different versions of the LHeC
and the FCC-he in ePb collisions, taken from \cite{Armesto}. Thin lines indicate, for pPb collisions at the LHC and the FCC, the kinematic coverage for rapidities 1,2,$\dots$,6 from right to left.}
\label{fig:kinematics}
\end{center}
\end{figure}

Ultraperipheral collisions (UPC) in which hadrons and heavy nuclei accelerated at very high energies generate a huge flux of quasi-real photons,  can be used to study high-energy $\gamma$$\gamma$,
$\gamma$p and $\gamma$A processes when the impact parameter is large enough to neglect hadronic interactions. Specifically, $\gamma$A collisions offer the possibility, through exclusive $J/\psi$ and $\Upsilon$ production, of constraining nPDFs in uncharted regions of the $x-Q^2$ plane, see the accessible kinematic region in Fig.~\ref{fig:kinematics} left. In this respect, the complementarity  with future high-energy electron-ion colliders, the Large Hadron Electron Collider (LHeC) and the FCC-he, and pA collisions at the FCC-hh, see Fig.~\ref{fig:kinematics} right, should be exploited.

\vskip -3mm








\end{document}